\documentclass[a4paper]{jpconf}
\usepackage{float} 
\usepackage{xurl} 
\usepackage{graphicx}
\usepackage[export]{adjustbox}
\begin{document}
\title{Software Training in High Energy Physics}




\author{Michel H.\ Villanueva $^1$, Sudhir Malik $^2$, Meirin Oan~Evans$^3$}

\address{$^1$Deutsches Elektronen-Synchrotron (DESY), Notkestraße 85, 22607 Hamburg, Germany }
\address{$^2$Physics Department, University of Puerto Rico Mayagüez, PR 00682, USA}
\address{$^3$Department of Physics and Astronomy, University of Sussex, Brighton BN1 9QH, UK}
\ead{sudhir.malik@upr.edu}

\begin{abstract}
Among the upgrades in current high energy physics (HEP) experiments and the new facilities coming online, solving software challenges has become integral for the success of the collaborations, The demand for human resources highly-skilled in both HEP and software domains is increasing. With a highly distributed environment in human resources, the sustainability of the HEP ecosystem requires a continuous effort in the equipment of physicists with the required abilities in software development. In this paper, the collective software training program in HEP and its activities led by the HEP Software Foundation (HSF) and the Institute for Research and Innovation in Software in HEP (IRIS-HEP) are presented. Experiment-agnostic, open, and accessible modules for training have been developed, focusing on common software material with ranges from core software skills needed by everyone to advanced training required to produce high-quality sustainable software. A basic software curriculum was built, and an introductory software training event has been prepared to serve HEP entrants. This program serves individuals with transferable skills that are becoming increasingly important to careers in the realm of software and computing, whether inside or outside HEP.
\end{abstract}

\section{Introduction}
Software skills are an integral part of the toolkit of any successful HEP experimentalist and maximizing the science from the hardware investments in current and future HEP projects rely critically on them. It is also a skill set that is transferable in case of non-HEP career evolution of people trained in HEP. Though software training is now key in many research fields, most users learn software skills only after joining a research program. Individual universities do not uniformly provide such training to students, prior to their beginning their Ph.D. research. Many domain-specific aspects add challenges to the learning process. Embarking on a HEP-specific path presents its own experiment-specific software environment challenges. The HEP community must play a role in bringing together a focus on these efforts to foray into sustainability and scalability and initiate learning early in the process of preparing a software-equipped future particle physicist. No one size fits all while imparting software training. A possible solution in HEP is to exploit our large-scale community structure and organize training within our research domains. Efforts like HEP Software Foundation (HSF)~\cite{HSF_homepage}, IRIS-HEP~\cite{IRISHEP_homepage},  FIRST-HEP~\cite{first_hep} and SIDIS
~\cite{sidis_cern} have taken strong and effective steps in this direction and impart training~\cite{training_springer} to those within the field and ancillary fields. 

\section{Training for Software in HEP}
\begin{figure}[h]
\begin{center}
\includegraphics[width=20pc]{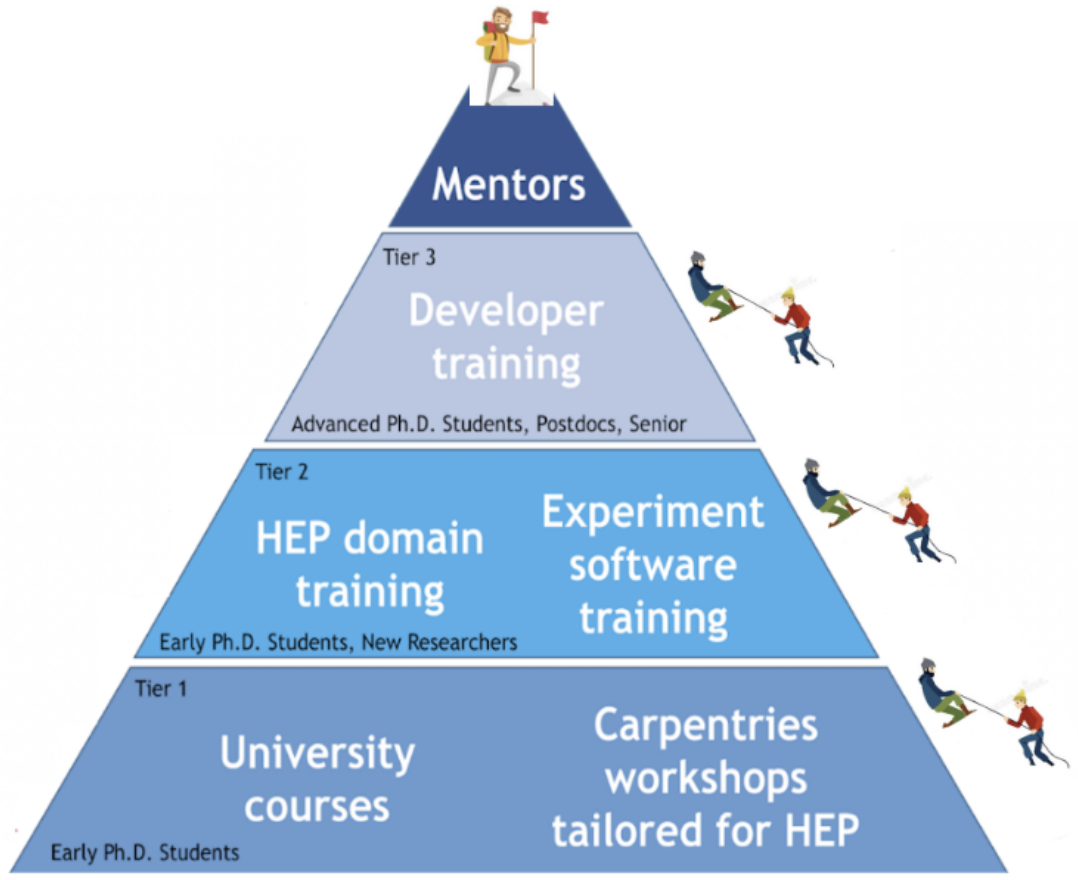}\hspace{5pc}%
\begin{minipage}[b]{25pc}\caption{\label{fig:pyramid}Evolution of HEP Education and Training}
\end{minipage}
\end{center}
\end{figure}

Trained software developers are critical resources in the success of current and future HEP experiments. It is therefore essential that the HEP community as a whole design and develop tools, methods, and resources to advance software skills among its community, that in turn will lead to solutions to software challenges. Developing a HEP workforce early is required to address the emerging needs, and unresolved bottlenecks in software and computing in scientific and engineering research workforce development, in general and HEP in particular. Undergrads, Master and PhD students may leave the field of HEP after completing their degrees. Imparting software skills while they pursue their degrees can help jump start contribution to the experiment, as well as prepare for another STEM degree or a career outside or inside academia. If this group pursues HEP they come well prepared with software to contribute or learn more.  It is a win-win situation for readying the workforce. Postdocs have a longer time span in HEP, even if they decide to leave the field of HEP for different reasons. Scientists and Physicists who choose to stay in the field always need to update their skills with new tools that disrupt the field, like Machine Learning. As shown in Figure \ref{fig:pyramid} the HSF/IRIS-HEP vision of evolution of training spans from early graduate students to mentors and has extended this fabric to undergraduate and master student level. Furthermore, the training outreach activities are reaching STEM aspirants in K-12 school students and their teachers. 

HEP software training is a cross-experiment, multi-domain activity and poses its own challenges. Each experiment has a big chunk of its own specific software. However, many HEP tools like ROOT and GEANT, coding languages like Python and C++, Operating Systems like Linux, version control systems like \emph{Github}, and algorithms like Machine Learning are common across the experiments. Some of the common HEP tools are common to the entire software community across all STEM and Computing disciplines. Some of our training and outreach events can be found here~\cite{training_hsf,training_irishep}. A rough estimate shows that we have about ~10K people in our community to train every year, ranging from Undergrads to Scientists. This implies that training activities need to scale and sustain with time so that the benefit reaches all and quickly. What adds to this challenge is that software training is a completely voluntary activity for mentors and participants. This requires a continuous motivation of new participants or experienced users to volunteer as mentors. 

\section{Community Building}
\begin{figure}[h]
\begin{center}
\includegraphics[width=35pc,frame]{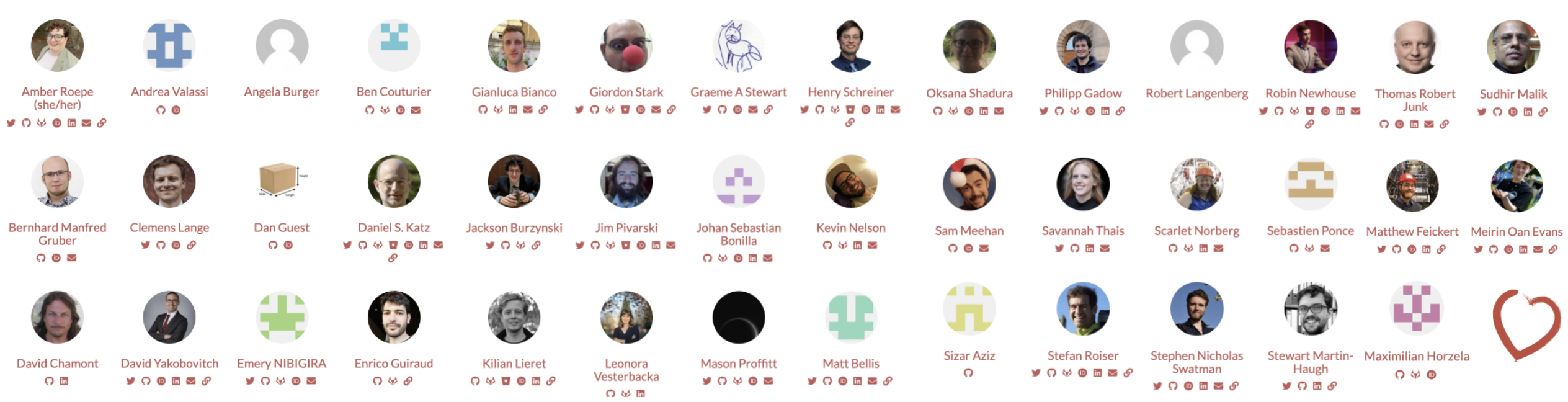}\hspace{5pc}%
\begin{minipage}[b]{28pc}\caption{\label{fig:community}Community of Mentors}
\end{minipage}
\end{center}
\end{figure}

Community building is central to HEP-wide software training. It is essential to build and improve training material, incorporate new ideas, make connections and acquire new mentors for sustainability and scalability. The HSF/IRIS-HEP training community of mentors is budding~\cite{hsf-community,hsf-community-building} and is voluntary-based, reflecting the vision that participants become mentors at different levels in the pyramid and feedback into the system. It has also expanded the mentor community to related fields like Nuclear Physics and Computing.  However, keeping the community growing and engaged needs constant nurturing, coordination, motivation, and involvement of several key and senior people, especially, from HEP labs and institutions. The mentors should also be able to see a value in continuing to remain a part of the community, at least for some reasonable period of time. The mentors' profiles (see a snapshot in Figure \ref{fig:community}) are advertised on our web page, that also shows links to their further professional details. 

\section{Training Content}
\begin{figure}[h]
\begin{center}
\includegraphics[width=30pc,frame]{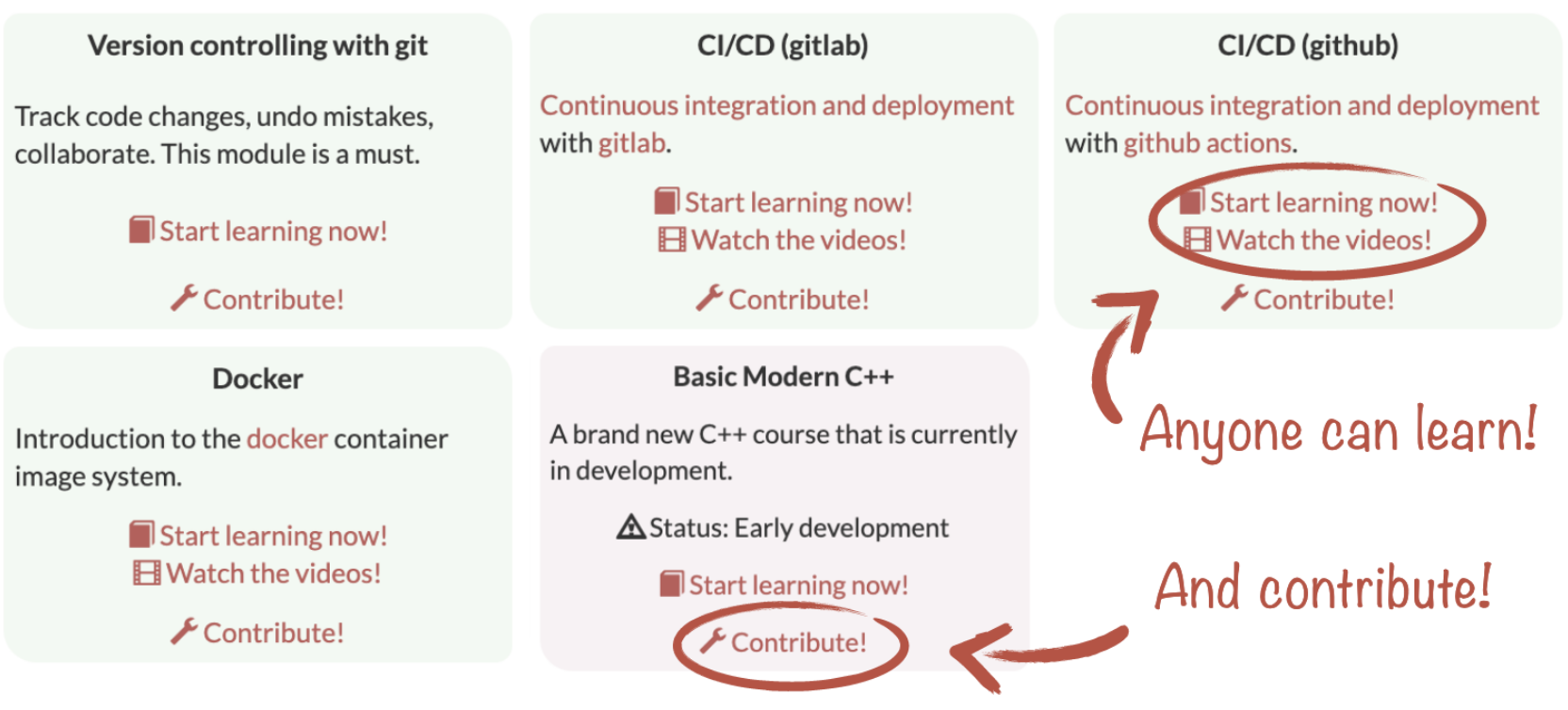}\hspace{5pc}%
\begin{minipage}[b]{28pc}\caption{\label{fig:curriculum}HEP Software Curriculum}
\end{minipage}
\end{center}
\end{figure}

Most HEP experiments have similar basic prerequisites that are actually common with other related areas like Nuclear Physics or Computing. We are building a training curriculum that consists of independent modules. This is necessitated by a view that students should be able to prioritize certain skills before others, as in many cases like HEP Physics Analysis it is expected to jump-start directly towards scientific results, with minimal time given for acquiring software knowledge or best practices. The current spectrum of modules consists of a basic skill set (\emph{Unix, Shell, Python, Git}) that serves newcomers, software engineering topics (Continuous Integration) with some specific to HEP data analysis and techniques, to advance topics (Parallel Programming and Machine Learning). All Modules are open-source, which is in line with our view of Open Science. Figure \ref{fig:curriculum} shows a snapshot of our training curriculum \cite{HSF-training-center}.

We have organized training events~\cite{HSF-training-events-url} in the last 3-4 years and trained about 1500 people. Due to COVID-19, virtual training events were held in 2020. Some advantages and disadvantages of in-person and virtual training, based on our training experience, are described in~\cite{training_springer}. Software Carpentry events populated with a basic skill set and some HEP specific training topics (like \emph{PyROOT, Uproot}) are being organized 2-3 times per year, supported by our community along with \emph{The Carpentries}~\cite{carpentries}, with which we currently have a membership, supported by IRIS-HEP. As our community and training topics mature, we aim to scale up the frequency of training on different topics.

\section{Impact of Training}
\begin{figure}[h]
\begin{center}
\includegraphics[width=35pc]{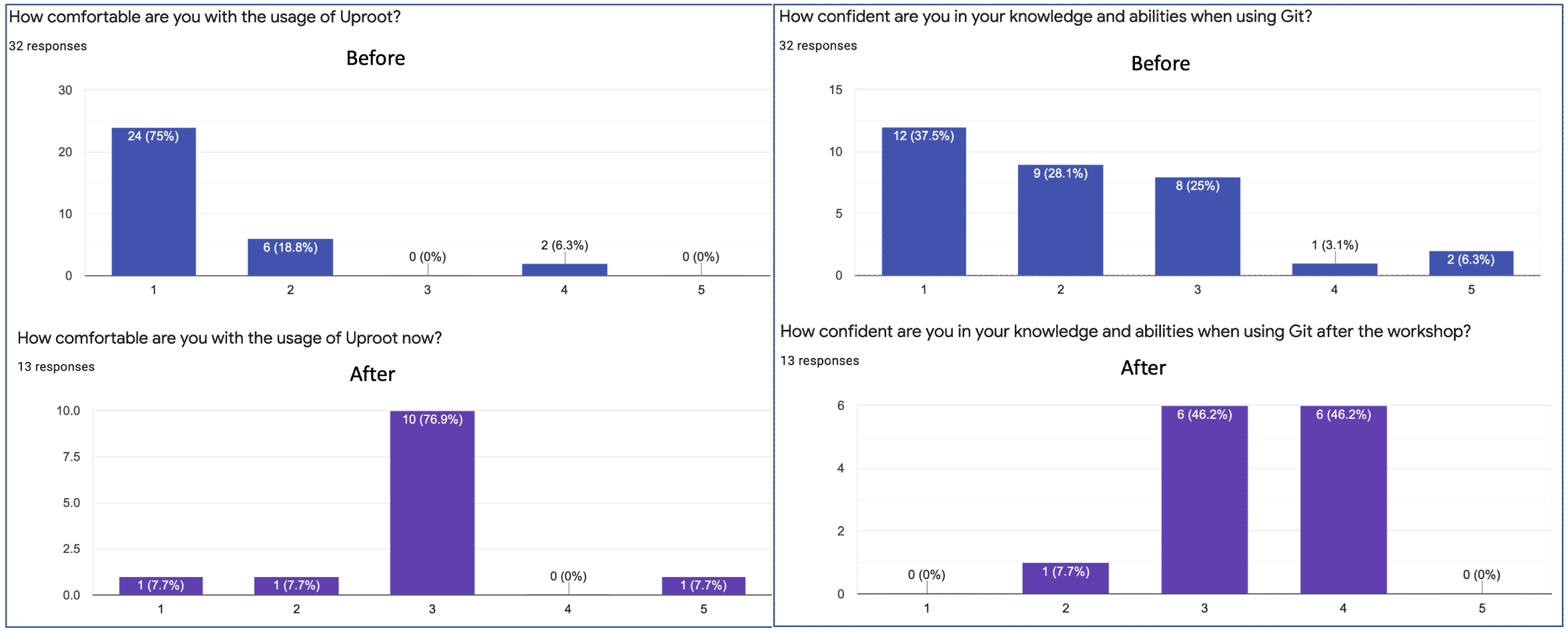}\hspace{5pc}%
\begin{minipage}[b]{28pc}\caption{\label{fig:survey}Survey Feedback for \emph{Uproot} and \emph{Git} modules }
\end{minipage}
\end{center}
\end{figure}

Assessing the impact of training is essential to its continued effectiveness, success and future. Every training format involves a pre-training and post-training survey. While the pre-training survey includes demographic and "How much do you already know" questions, the post-training is focused on what has been learnt and compares the skills outcome before and after the training. One such example is shown in Figure \ref{fig:survey}, where a comparison is made of knowledge in \emph{Uproot} and \emph{Git} modules before and after the training. Note the difference in the statistics of people who took the survey before and after the training. Getting everyone to take surveys, especially after the training, has been a challenge ad we are trying to address that. However, it is clear that the training is transforming the skill-set of participants. The number of people who know more confidently about the module moves from left (before the training) to the right (after the training) for both the example modules shown.

\section{The Software Training Challenge}
One of the key things to succeed in software training is to have an up-to-date website~\cite{hsf_trainingurl}. As a training community, regular brainstorming sessions are organised every few months to have experts re-visit our training material and surrounding pedagogy. In addition, there are weekly meetings to discuss progress organising events and developing materials~\cite{hsf_trainingmeetings}. A rough estimate shows a target HEP community of around ten thousand Postdocs, PhD students and Undergrads that work on HEP projects in any given year. There are similar people from related fields like Nuclear Physics, Astrophysics and Computer Science who we constantly try to reach out to. There are also scientists and faculties who may want to learn or update their software skills. Our goal is that every HEP graduate student should attend the basic training to the very least. This translates into an increase in the frequency of our events and size in the community of individuals - facilitators, participants, instructors, experts and hosts. In the recent past we have succeeded in expanding collaboration with related communities like Nuclear Physics and Neutrino. We want to explore ways to incentivize new trainers to continually contribute. We are especially committed~\cite{hsf_diversity} to Diversity and Inclusion where everyone feels welcome to participate. We also host Outreach events to promote early software awareness and skill development among high school students. Some of our events can be accessed via~\cite{hsf_outreach}.

Most of our training events have been in the COVID-19 era. As we come out of it, we aim to look into more creative ways to scale and sustain. Evaluating curriculum receptivity, building a course around basic curriculum for HEP beginners, giving course credits and certificates of participation as an incentive, are some of these ways. In the long term, having a financial model that builds regional and local training capacity is essential to sustain and scale. Another key part of this model is to establish equity, diversity, inclusion, and accessibility. This participation will span across HEP communities, under-resourced institutions and communities in different geographical regions. Opportunities to grow professionally and have career paths for the core team and volunteers are essential as we plan for the future.

\section{Conclusions}
We discussed the key feature of the HSF/IRIS-HEP software training program that aims at ensuring sustainability of software in HEP for years to come. Our training material allows open-source access. This process enables continual feedback to improve the curriculum. The program promotes a culture within HEP of valuing software skills and teaching of those skills to others. A growing community with a special focus on inclusion and diversity in science is essential to sustain and scale the program. 

We thank NSF grants OAC-1836650,  OAC-1829707 and OAC-1829729 for support of the training program.

\section*{References}
\bibliography{iopart-num}


\end{document}